\documentclass[pra,aps,twocolumn]{revtex4}

\usepackage{amsfonts}
\usepackage{amssymb}
\usepackage[dvips]{graphicx}

\usepackage[T1]{fontenc}
\usepackage[latin2]{inputenc}

\usepackage{amsmath}

\usepackage{mathbbol,bbm}
\usepackage{epic}

\newcommand{\Cx}{{\mathbb C}}

\newcommand{\idty}{\Eins}
\DeclareMathOperator{\id}{id}

\DeclareMathOperator{\tr}{Tr}
\newcommand{\<}{\langle}
\renewcommand{\>}{\rangle}

\renewcommand{\c}[1]{\mathcal{#1}}

\renewcommand{\r}[1]{\mathrm{#1}}

\def\zcal{\cal Z}

\def\duzomniejsze{<\kern-.7mm<}
\def\duzowieksze{>\kern-.7mm>}

\def\textbf#1{{\bf #1}}
\def\beq{\begin{equation}}
\def\eeq{\end{equation}}
\def\be{\begin{equation}}
\def\ee{\end{equation}}
\def\ben{\begin{eqnarray}}
\def\een{\end{eqnarray}}
\def\beqa{\begin{eqnarray}}
\def\eeqa{\end{eqnarray}}
\def\eea{\end{array}}
\def\bea{\begin{array}}
\newcommand{\bei}{\begin{itemize}}
\newcommand{\eei}{\end{itemize}}
\newcommand{\bee}{\begin{enumerate}}
\newcommand{\eee}{\end{enumerate}}

\def\Tau{\tcal}
\def\hcal{{\cal H}}
\def\tcal{{\cal T}}
\def\lcal{{\cal L}}
\def\lscal{\lambda}

\def\ccal{{\cal C}}

\def\kcal{{\cal K}}

\def\tr{{\rm Tr}}
\def\id{{\rm I}}

\def\>{\rangle}
\def\<{\langle}
\def\blacksquare{\vrule height 4pt width 3pt depth2pt}

\def\ot{\otimes}

\def\dt#1{{{\kern -.0mm\rm d}}#1\,}

\def\loops{\kcal}

\def\sigalpe{{\sigma_\alpha'}^{\kern-.7mm E}}
\def\sigalpb{{\sigma_\alpha'}^{\kern-.7mm B}}

\newtheorem{lemma}{Lemma}

\newtheorem{theorem}{Theorem}
\newtheorem{proposition}{Proposition}
\newtheorem{definition}{Definition}
\newtheorem{fact}{Fact}

\def\bep{\begin{proposition}}
\def\eep{\end{proposition}}
\def\bel{\begin{lemma}}
\def\eel{\end{lemma}}

\def\bet{\begin{theorem}}
\def\eet{\end{theorem}}
\def\bed{\begin{definition}}
\def\eed{\end{definition}}
\def\bef{\begin{fact}}
\def\eef{\end{fact}}

\begin{document}


\author{R. Alicki$^{1,2}$}
\author{M. Horodecki$^{1,2}$}
\author{P. Horodecki$^{1,3}$}
\author{R. Horodecki$^{1,2}$}

\affiliation{
$^1$ National Quantum Information Centre of Gda\'nsk, Poland \\
$^2$ Institute of Theoretical Physics and Astrophysics, University of Gda\'nsk, Poland \\
$^3$ Faculty of Applied Physics and Mathematics, Gda\'nsk University of Technology, Poland
}

\title{On thermal stability of topological qubit in Kitaev's 4D model}

\begin{abstract}
We analyse stability of the  four-dimensional
Kitaev model - a candidate for scalable quantum memory - in finite 
temperature within the weak coupling Markovian limit.  It is shown 
that, below a critical temperature, certain topological qubit 
observables $X$ and  $Z$  possess relaxation times exponentially long in the size 
of the system. Their construction involves polynomial
in system's size algorithm which uses as an input the results of 
measurements performed on all individual spins. We also discuss 
the drawbacks of such candidate for quantum memory and mention the implications of the stability of qubit for 
statistical mechanics.
\end{abstract}

\maketitle

\section{Introduction}
While quantum computation offers algorithms which can outperform the classical ones, they 
are very fragile with respect to external disturbance. Therefore, along with the discoveries of 
fast algorithms, the question of how to protect quantum computation against decoherence was 
the subject of extensive studies. As a result the whole domain was created known as fault tolerant quantum 
computation \cite{Nielsen-Chuang}. The famous {\it threshold theorems} \cite{AharonovBO1996-FT,ZurekFT}, 
saying that arbitrary long quantum 
computation is  possible provided the error per gate is below certain threshold has given the hope, that it is possible in principle to overcome the decoherence. However the initial theorems are based on phenomenological model of noise, and the problem,  has not been solved within Hamiltonian dynamics \cite{AHHH2001,TerhalB-FT,Alicki2004-FT-comment,AharonovKP-FT-NM}. Even the problem of whether one 
can store qubits is open. 

There is, though a class of candidates for quantum memories, which are in between realistic description and the phenomenological one: the Kitaev models of topological quantum memory \cite{KitajewPreskill-FTmem,KitaevFT,BombinMD2006-top-branyons}. There is a heuristic reasoning, according to which such memories are instable in two dimensions \cite{KitajewPreskill-FTmem,AlickiH2006-drive}, and stable in four dimensions (similarly like Ising model represent a  stable classical memory in 2D, but not in 1D) \cite{KitajewPreskill-FTmem}. Behaviour of 
of Kitaev models in finite temperature was then investigated 
(see e.g. \cite{AlickiFH2007-Kitaev,NussinovO2007-topord1,NussinovO2007-topord2,
Kay2008-K2D,IblisdirGAP2008-toptemp}). Quite recently 
the thermal instability of 2D model has been rigorously proved in \cite{AlickiFH2008-K2D}. In the present paper, 
we deal with the 4D Kitaev's model of  Ref. \cite{KitajewPreskill-FTmem} and prove 
rigorously, within  Markovian weak coupling approximation,  that the model provides thermally stable  
qubit. To this end we  use the formalism of quantum semigroup theory \cite{Alicki-Lendi2}, which has 
been successfully applied to analysis of Kitaev 2D model in Ref. \cite{AlickiFH2008-K2D}. 
As a byproduct we obtain a very useful general upper bound for decay rate. 
We perform our analysis in parallel for 3D and 4D case. Indeed, though in 3D case 
only one of the qubit observables is stable, as argued in \cite{KitajewPreskill-FTmem}, 
it is much more transparent and the reasoning is the same as in 4D case.   
Since the very stability of qubit is not sufficient for a good quantum memory, 
we also discuss the  open problems concerning existence of self-correcting quantum memory. Implications 
for description of thermodynamical limit are also discussed. 

The paper is organized as follows. In section \ref{sec:Markov} we provide some basic notions and results
concerning Markovian weak coupling limit. We show, in particular, how the rate of decay expressed in 
terms of noise generator is related  to fidelity criterion. Finally we provide a general upper bound 
for decay rate. In section \ref{sec:q-cl} we show that analysis of noisy evolution 
of some particular topological observables is reduced to the study of a classical model. 
Next  (sec. \ref{sec:stab-Kitaev}) we provide conditions for stability of these observables in terms of one-step autocorrelation functions. In  sec. \ref{sec:top-stab} we finally prove 
the stability of the observables. In sec. \ref{sec:alg}
we provide polynomial algorithm to measure the observables. Finally (sec. \ref{sec:conc}) we discuss 
remaining open problems for existence of self-correcting quantum memory, as well as importance of 
the result for description of systems in thermodynamical limit.

\section{Markovian approximation in weak coupling limit}
\label{sec:Markov}
Let us first we briefly sketch the
general setup and properties of Davies generators. A quantum system with discrete energy spectrum
is coupled to a collection of heat baths leading to the global Hamiltonian
\begin{equation}
 H = H^{\r{sys}} + H^{\r{bath}} + H^{\r{int}}
 \qquad\text{with}\qquad
 H^{\r{int}} = \sum_\alpha S_\alpha \otimes f_\alpha, 
\end{equation}
where the $S_\alpha$ are system operators and the $f_\alpha$ bath
operators. The main ingredients are the Fourier transforms $\hat
h_\alpha$ of the autocorrelation functions of the $f_\alpha$. The
function $\hat h_\alpha$ describes the rate at which the coupling is
able to transfer an energy $\omega$ from the bath to the
system. Often a minimal coupling to the bath is chosen, minimal in the
sense that the interaction part of the Hamiltonian is as simple as
possible but still addresses all energy levels of the system
Hamiltonian in order to produce finally an ergodic reduced dynamics.
The necessary and sufficient condition for ergodicity is~\cite{Spohn1977-ergod,Frigerio1978-ergod}
\begin{equation}
 \bigl\{ S_\alpha, H^{\r{sys}} \bigr\}' = \Cx\, \idty,
\end{equation}
i.e.\ no system operator apart from the multiples of the identity
commutes with all the $S_\alpha$ and $H^{\r{sys}}$. 

We begin by introducing the Fourier decompositions of the $S_\alpha$'s as they
evolve in time under the system evolution
\begin{equation}
 \r e^{itH^{\r{sys}}}\, S_\alpha\, \r e^{-itH^{\r{sys}}} 
 = \sum_\omega S_\alpha(\omega)\, \r e^{-i\omega t}. 
\end{equation}
Here the $\omega$ are the Bohr frequencies of the system Hamiltonian. From
self-adjointness we have the relation
\begin{equation}
 S_\alpha(-\omega) = S_\alpha(\omega)^\dagger.
\end{equation}
The weak coupling limit procedure then returns the following equation
for the evolution of the spin system in Heisenberg picture
\begin{align}
 \frac{dX}{dt} 
 &= i[H^{\r{sys}},X] + \c L_{\r{dis}}(X) =: \c L(X) 
\\
 \c L_{\r{dis}}(X)
 &= \frac{1}{2} \sum_\alpha \sum_\omega \hat h_\alpha(\omega) \Bigl(
 S_\alpha^\dagger(\omega)\, [X,S_\alpha(\omega)] + 
 \\
 &+ [S_\alpha^\dagger(\omega),X]\,
 S_\alpha(\omega) \Bigr)
\label{gen}
\end{align}
For thermal baths one has moreover the relation
\begin{equation}
 \hat h_\alpha(-\omega) = \r e^{-\beta\omega}\, \hat h_\alpha(\omega)
\end{equation} 
which is a consequence of the KMS~condition \cite{Alicki-Lendi2}. The operator $\c L$
generates a semigroup of completely positive identity preserving
transformations of the spin system.
It describes the reduced dynamics
in the Markovian approximation and enjoys the following properties
\begin{itemize}
\item
 The canonical Gibbs state with density matrix 
 \begin{equation}
  \rho_\beta = \frac{\r e^{-\beta H^{\r{sys}}}}{\tr \Bigl( \r e^{-\beta
  H^{\r{sys}}} \Bigr)}
 \end{equation}
 is a stationary state for the semigroup, i.e.\
 \begin{equation}
  \tr \Bigl( \rho_\beta\, \r e^{t\c L}(X) \Bigr) = \tr \bigl(
  \rho_\beta\, X \bigr).
 \end{equation}
\item
 The semigroup is relaxing, i.e. for any initial state $\rho$
 of the system
 \begin{equation}
  \lim_{t\to\infty}\ \tr \Bigl( \rho\, \r e^{t\c L}(X) \Bigr) = \tr \bigl(
  \rho_\beta\, X \bigr).
 \end{equation}
\item
 Furthermore, the generator satisfies the detailed balance condition,
 often called reversibility. Writing $\delta(X) :=
 [H^{\r{sys}},X]$,
\begin{equation}
 [\delta, \c L_{\r{dis}}] = 0
\end{equation}
and
\begin{equation}
 \tr \Bigl(\rho_\beta\, Y^\dagger\, \c L_{\r{dis}}(X) \Bigr) 
 = \tr \Bigl(\rho_\beta\, \bigl(\c L_{\r{dis}}(Y)\bigr)^\dagger\, X
 \Bigr).
\end{equation}
The last equation expresses the self-adjointness of the generator with
respect to the scalar product defined by the equilibrium state. The
space of observables equipped with the scalar product
\begin{equation}
 \<X,Y\>_\beta := \tr \rho_\beta\, X^\dagger\, Y 
\end{equation}
is called the Liouville space and the generator of the reduced dynamics
is a normal matrix on that space, i.e.\ the Hermitian and
skew-Hermitian parts of the generator commute.
\end{itemize}
Finally it is known that $-\lcal$ is a positive operator,
hence  it has nonnegative eigenvalues. Moreover 
$L(I)=0$, and for ergodic systems eigenvalue $0$ is nondegenerate. 

\subsection{Autocorrelation functions, decay rate and fidelity}
Suppose that for observable $X$ satisfying $\<X,X\>_\beta=1$
$\<X,I\>_\beta=0$ we have 
\be
-\< X, \lcal (X)\>_\beta \leq \epsilon.
\ee
Then the autocorrelation function of the observable satisfies
\be
\<X, e^{\lcal t}X\>_\beta\geq   e^{-\epsilon t}
\label{eq:X-rate}
\ee
One proves it easily, by decomposing $X$ into 
normalized eigenvectors of $\lcal$, and using convexity of function 
$e^{-x}$. Thus to show that an observable $X$ is stable, 
it is enough to estimate $-\<X,\lcal(X)\>$, which can be therefore called {\it decay rate} for the observable $X$.
If this quantity  decreases exponentially with size of the
system, we obtain stability. 

Let us now rephrase it in the language of fidelity. Namely, suppose we have 
observables $X$ and $Z$ satisfy commutation rules of Pauli algebra. 
They generate subalgebra which defines a virtual qubit, the one to be protected.  Let the  induced 
tensor product  on the total Hilbert space be 
\be
\hcal=\hcal_Q\ot \hcal_{anc}.
\ee
Now, we fix some state $\rho_{anc}$ on the system $\hcal_{anc}$. For any state $\psi$ 
of qubit the initial state of the total system is $\rho_{Q,anc}(0)=|\psi\>\<\psi|\ot\rho_{anc}$.
Then the system evolves into state $\rho_{Q,anc}(t)$, and finally, 
the ancilla is traced out. Thus the fidelity is given by 
\be
F(\psi)=\<\psi|\rho^{out}_Q(t)|\psi\>
\ee
where 
\be
\rho^{out}_Q(t)=\tr_{anc}(\rho_{Q,anc}(t)).
\ee
Let us denote the fidelity averaged uniformly over the states of qubit by $\overline F$.
\begin{proposition}
\label{prop:fidelity}
With the above notation, suppose now that the Gibbs state is of the form 
\be
\rho_\beta=\frac12 \id_Q\ot\rho_{anc},
\label{prop:gibbs-prod}
\ee
where $\rho_{anc}$ is a state on ancilla. We then have 
\be
\overline F\geq \frac12(\<X, e^{\lcal t}X\>_\beta+\<Z, e^{\lcal t}Z\>_\beta)\geq e^{-\epsilon t}
\ee
where $\epsilon$ is upper bound for the rates $-\<X,\lcal (X)\>_\beta$ and $-\<Z,\lcal(Z)\>_\beta$.
\end{proposition}

{\bf Proof.} 
Let $F_x$ be given by
\be
F_x=\frac12(F(|0\>)+F(|1\>))
\ee
where $|0\>, |1\>$ are eigenstates of $X$ treated as observable on system $Q$. Similarly we define $F_z$.
Using results of \cite{Hofmann2004-fidelity} and \cite{gentele} one finds that 
\be
\overline F\geq F_x + F_z -1
\ee
Thus it is enough to estimate e.g. $F_x$. Using the property (\ref{prop:gibbs-prod}) 
and orthogonality $X\perp I$ one finds that 
\be
F_x=\frac12(1+\<X,e^{\lcal t}X\>_\beta).
\ee
Combining the last two formulas with (\ref{eq:X-rate}) ends the proof.

\subsection{Upper bound for decay rate}
We now present a useful bound for decay rate, which holds for 
operators $X$ which are eigenvectors 
of $[H,\cdot]$. For such operators, one computes 
\begin{align}
&-\<X, \lcal(X)\>_\beta = \sum_{\omega\geq 0} \hat h (\omega) 
\Bigl( \<[S_\alpha(\omega), X],[S_\alpha(\omega), X]\>_\beta \nonumber \\
&+e^{-\omega \beta}
\<[S_\alpha(-\omega), X],[S_\alpha(-\omega), X]\>_\beta
\Bigr)\leq  \\
&\leq 2\sum_\omega \hat h(\omega)
\<[S_\alpha(\omega), X],[S_\alpha(\omega), X]\>_\beta \leq \nonumber \\
&\leq 2\hat h_{\max} \sum_\omega
\<[S_\alpha(\omega), X],[S_\alpha(\omega), X]\>_\beta.
\end{align}
where 
\be
\hat h_{\max}=\sup_{\omega \geq 0} \hat h(\omega).
\label{eq:hmax}
\ee
Since $X$ and 
$S_\alpha(\omega)$ 
are eigenvectors of $[H,\cdot]$ it follows that 
$[S_\alpha(\omega),X]$ are eigenvectors  of 
$[H,\cdot]$ too, hence they are mutually orthogonal. We thus can write
\be
\sum_\omega
\<[S_\alpha(\omega), X],[S_\alpha(\omega), X]\>_\beta
= \sum_{\omega,\omega'}
\<[S_\alpha(\omega), X],[S_\alpha(\omega'), X]\>_\beta.
\ee
However from definition of $S_\alpha(\omega)$ 
it follows that 
\be
\sum_\omega S_\alpha(\omega)=S_\alpha
\ee
This gives 
\be
-\<X,\lcal(X)\>_\beta \leq 2 \hat h_{\max} \sum_\alpha 
\<[S_\alpha, X],[S_\alpha, X]\>_\beta
\label{eq:przerwa}
\ee 
The advantage of the formula is that the only place where the self-Hamiltonian 
appears is the Gibbs state in scalar product.

\section{From quantum to classical 
in Kitaev-type models}
\label{sec:q-cl}
We consider a system of $N$ spin-$1/2$ systems. 
For any set $S$ of spins let us denote $X_S=\Pi_{j\in S} \sigma^x_j$, $Z_S=\Pi_{j\in S} \sigma^z_j$.
Consider now Hamiltonian of the form
\be
H=-\sum_s X_s -\sum_c Z_c
\ee
and we assume that the sets $s$ and the sets $c$ are chosen in such a way that the operators 
$X_s$ and $Z_c$ commute with each other. 
Consider also the following coupling to environment
\be
H_{int}=\sum_j \sigma_j^x \ot f_j+\sum_j \sigma_j^z\ot \tilde f_j.
\ee
Then Davies operators fall into two types:
\ben
\label{eq:dav-a}
&&a_\alpha=\sigma_x^j P_\alpha \\
\label{eq:dav-b}
&&b_\alpha=\sigma_z^j R_\alpha
\een
where $P_\alpha$ belong to algebra spanned by 
those operators $Z_c$ whose support contains the $j$-th spin
and $R_\alpha$ belongs to algebra spanned by operators $X_s$,
whose support contains $j$-th spin. (If the spin 
does not belong to support of any $s$, then $P_\alpha=I$,
and similarly for $R$. However,
in Kitaev-type models this latter case does not occur). The dissipative generator has the following form 
\be
\lcal=\lcal_x + \lcal_z, 
\ee
where $\lcal_x,\lcal_z$ consist of Davies operators of type $a$ 
and $b$ respectively.
The Davies operators describe the elementary 
noise processes. In 2D model, they are creation, anihilation and motion of two types of point-like anyons. 
In 4D model, excitations are not point-like, and the processes are creation, anihilation and 
two types of modification of loops (see \cite{KitajewPreskill-FTmem}, sec. X, and secs. \ref{sec:top-stab}, \ref{subsec:XZ-Kitaev} of  the present paper).

Consider now observables of the form $X_S$ and $Z_T$,
where $S,T$ are some subsets of spins. 
Let us assume that $X_S$ and $Z_T$ commute with 
all $X_s$ and $Z_c$. Then $X_S$ commutes 
with Davies operators of type $a$ and $Z_T$ commutes with 
Davies operators of type $b$. 
Therefore from \eqref{gen} we get that 
\be
\lcal(X_S)= \lcal_z(X_S),\quad \lcal(Z_T)=\lcal_x(Z_T)
\ee

Consider now a modification of the model. Let the Hamiltonian 
be of the form
\be
H=-\sum_s X_s
\ee
and the coupling with environment be of the form
\be
H_{int}= \sum_j\sigma_z^j \ot \tilde f_j.
\ee
Then dissipative generator for this model consists of 
Davies operators \eqref{eq:dav-b} i.e. it is 
given just by $\lcal_z$.
We obtain

\begin{proposition}
\label{prop:cl-model}
Let $X_s=\Pi_{j\in s}\sigma^x_j$, $Z_c=\Pi_{j\in c}\sigma^z_j$ 
where the sets $s$ and $c$ are chosen in such way that 
$X_s$ and $Z_c$ commute for all $s,c$. 
Consider $X_S$ which commutes with all $X_s$ 
and $Z_c$.
Then 
\be
\lcal(X_S)=\lcal'(X_S)
\ee
where $\lcal$ is dissipative generator coming from 
\be
H=-\sum_s X_s -\sum_c Z_c,\quad 
H_{int}=\sum_j \sigma_j^x \ot f_j+\sum_j \sigma_j^z\ot \tilde f_j.
\ee
and $\lcal'$ is dissipative generator coming from 
\be
H'=-\sum_s X_s,\quad
H'_{int}=\sum_j \sigma_j^z\ot \tilde f_j.
\ee
Moreover 
\be
\tr \bigl(\rho_\beta X^\dagger \lcal(X)\bigr)=\tr \bigl(\rho'_\beta X^\dagger \lcal'(X)\bigr),
\ee
where  $\rho_\beta=\frac{1}{\zcal}e^{-\beta H}$ and  
$\rho'_\beta=\frac{1}{\zcal'}e^{-\beta H'}$ respectively.
Analogous result holds for $Z_T$, which commutes with all $X_s$ 
and $Z_c$. 
\end{proposition}

{\bf Remark.} Further in text, $\<\cdot,\cdot\>_\beta$ will denote scalar product with 
the Gibbs state of type $\rho'_\beta$ (with suitable $H'$, depending whether we talk about $X$ 
or $Z$). 



\subsection{Observables $X$ and $Z$}
\label{subsec:XZ-Kitaev}

\subsubsection{$3D$ Kitaev model}

The Hamiltonian for 3D Kitaev model is given by \cite{KitajewPreskill-FTmem}
\be
H=-\sum_s X_s -\sum_{c} Z_c
\ee
where each $s$ denotes set of four plaquettes which share common 
link, and and each $c$ is six plaquettes forming cube. 
We will now define a class of  observables of interest.
To this end we will use 
observable $X_C$ with $C$ being set of 
parallel plaquettes forming a loop that winds around the torus
(there are three homologically inequivalent choices,
we will consider a fixed one of them).  Such observable is very unstable, hence we may call it 
"bare qubit observable". One needs to "dress" it with another dichotomic observable which would 
store the error syndrome. The latter observable 
will then belong to the abelian algebra spanned by star observables $X_s$, hence  
depending solely on atomic projectors of the algebra which correspond to 
configurations $\loops$  of excited links (stars can 
be labeled by the links - their centers). Let us call 
the projectors $P_{\loops}$. The needed observable will be thus of the form  
\be
F_x=\sum_{\loops} \lambda_\loops P_\loops,
\ee
where $\lambda_\loops=\pm1$. We shall not determine the values of $\lambda_\loops$ at the moment.  
They will emerge from our  analysis of stability in sec. \ref{sec:top-stab} and will be then described in 
sec. \ref{sec:alg}.

The full "dressed observable"  is the product $X_C F_x$. According to Proposition \ref{prop:cl-model} it evolves according to classical model with Hamiltonian 
\be
H_X=-\sum_s X_s
\ee
coupled to environment via operators $\sigma^z_j$. 
The model is known as $Z_2$ gauge Ising model (the Ising variables 
are in our case eigenvectors of $\sigma^x_j$) \cite{Kogut1979-gauge}. 

One can define analogous observable $Z_{P} F_z$.
However in $3D$ there will be no X-Z symmetry. 
The observable $Z_P$ is associated with plane,
and atomic projector of algebra spanned by $Z_c$ is labeled 
by configurations of points (i.e. the  plaquettes) rather than by loops. 
Observable $Z_P F_z$ is evolving according to 
the model with $H_Z=-\sum_c Z_c$ coupled via $\sigma_j^x$.
It will not be stable (as pointed out in \cite{KitajewPreskill-FTmem}) and most likely,
one can prove it by use of techniques worked out in \cite{AlickiFH2008-K2D}.

\subsubsection{$4D$ Kitaev model}
In four dimensional model  the spins again sit on  plaquettes, 
and  the Hamiltonian is similar as in 3D case:
\be
H=-\sum_s X_s - \sum_c Z_c
\ee
The only difference is that the star $s$ has six plaquettes, 
because there is six plaquettes common to a single link.
Thanks to it there is symmetry: We fix two 
planes $p_1$ and $p_2$ 
on the lattice and on the dual lattice, respectively,
obtaining  bare qubit observables $X_{p_1}$ and $Z_{p_2}$.
Then candidates for stable observables 
will be the dressed ones $X_{p_1} F_x$, $Z_{p_2} F_z$.
The latter will again evolve separately, 
and since $4D$ lattice is self-dual, 
the evolutions are the same. 
We arrive at the $4D$ $Z_2$ gauge Ising model.  

If we prove that e.g. observable of the form $X_{p_1} F_x$ 
is stable, then also similar $Z_{p_2} F_z$
will be stable too, so that we will obtain  
stable qubit.

\section{Stability conditions  for Kitaev model}
\label{sec:stab-Kitaev}
\subsection{Bound for decay rate for dressed observables}
The bound (\ref{eq:przerwa}) applied to generator consisting of Davies generators (\ref{eq:dav-a}), (\ref{eq:dav-b})
takes the form
\ben
&&-\<A,\lcal(A)\>_\beta \leq 2 \hat h_{\max} \sum_j\<[\sigma_j^x,A],[\sigma_j^x,A]\>_\beta +\nonumber\\
&&+ \sum_j\<[\sigma_j^z,A],[\sigma_j^z,A]\>_\beta. 
\een
The quantity $h_{\max}$ given by (\ref{eq:hmax}) is a constant independent of the size of the system. This is due to 
the fact that Kitaev models exhibits strong locality property, implying that there 
is a constant number of frequencies involved in the generator (e.g. just one positive frequency in 2D model)
which are independent of the number of spins $N$. 

Since the observables $Z=Z_PF_z$, $X=X_C F_x$ (or analogous ones from 4D model) commute with Hamilotnian, 
the bound is applicable. We obtain  
\begin{align}
-\<X,\lcal(X)\>_\beta \leq 
4 \hat h_{\max}  \sum_j(1 - \< X, \sigma_j^z X \sigma_j^z\>_\beta)
\nonumber \\
-\<Z,\lcal(Z)\>_\beta \leq 
4 \hat h_{\max}  \sum_j(1 - \< Z, \sigma_j^x Z \sigma_j^x\>_\beta)
\end{align}
where $j$ runs over all spins. We see that the problem of decay of time autocorrelation
function has been reduced to the much simpler problem of  "one step" autocorrelation function.

\subsection{Gibbs state is concentrated on 
configurations without long loops}

First we will estimate probability 
that a configuration has loop of length $l$.
We shall use the Peierls argument following
Dennis et al. \cite{KitajewPreskill-FTmem} and  Griffiths \cite{Griffiths1964-peierls}.
To this end we first estimate probability  that a fixed loop $\lscal$ with length $l$ emerges. 
Let $\ccal$ be the set of all configurations  which contain loop $\lscal$.
The probability is then given by
\be
P(\lscal)= 
{\sum_{\kcal\in \ccal}e^{-\beta E(\kcal)}\over 
\sum_\kcal e^{-\beta E(\kcal)}}
\ee
where in denominator we have sum over all configurations. 
For any configuration $\kcal$ containing $\lscal$ 
we flip spins on a chosen surface whose boundary is $\lscal$,
obtaining new configuration $\kcal^*$ 
which differs from $\kcal$ only in that the loop  $\lscal$
is not present anymore. Hence $E(\kcal)=E(\kcal^*) e^{-\beta l}$ (or the quantities here are taken 
to be dimensionless).
Thus we write 
\be
P(\lscal)= {
e^{-\beta l}\sum_{\kcal^*\in \ccal}e^{-\beta E(\kcal^*)}\over 
\sum_\kcal e^{-\beta E(\kcal)}}
\ee
Leaving in denominator only configurations $\kcal^*$, 
we can only decrease it, so that  $P(\lscal)\leq e^{-\beta  l}$.

Now, the probability $P(l)$ of appearing a configuration which has 
a loop of length $l$ is bounded by 
the number of all possible loops of length $l$ 
times $e^{-\beta l}$. A trivial bound for the number 
of loops in cube of linear size $L$ in dimension $d$,
that start from a fixed node is $2d(2d-1)^l$. 
This should be multiplied by the number of nodes, 
which is proportional to the volume i.e. 
polynomial in linear size $L$ of the system.
Finally, we obtain that 
\be
P(l)\leq poly(L) \mu^l  e^{-\beta l} = poly(L) 
e^{-l(\beta-\ln \mu)}
\ee 
where $\mu$ is a constant depending only on $d$. 
Thus below certain critical temperature $T_{\r crit}$ we have 
\be
P(l)\leq poly(L) e^{-\delta l }
\ee
where $\delta=\beta-\ln \mu$  is positive and does not depend on the size of the system. 
We then evaluate probability  of appearing 
a configuration that has a loop  greater  than $L'$ 
\be
P(l\geq L') \leq poly(L) \sum_{l=L'}^{\infty} e^{-\delta l}=
poly(L) e^{-\delta L'} {1\over 1- e^{-\delta}}
\ee
Thus we see that below $T_{\rm crit}$ the probability of obtaining 
e.g. a loop of length $L/8$ or greater is exponentially decaying in $L$. 

\subsection{Stability of Kitaev 4D model}
In next section we shall prove that for configurations having only loops shorter than $L'=L/8$ a single flip does not 
change observables $X$ and $Z$ for Kitaev 4D model. This implies that 
\be
\sum_j(1- \< Z, \sigma_j^x Z \sigma_j^x\>_\beta) \leq \sum_j 2 P(l\geq L')
\ee  
so that 
\be
-\<Z,\lcal(Z)\>\leq {\rm poly}(L) e^{-\delta' L}
\ee
where $\delta'=\delta/8$ is a constant that is positive below some critical temperature. The same happens 
for observable $X$, hence due to proposition \ref{prop:fidelity} the decay time of fidelity 
is exponentially long in size of the system. 

\section{Stability of topological observables}
\label{sec:top-stab}
In previous section we have shown that below certain critical temperature $T_{\rm crit}$ the 
Gibbs state is concentrated on configurations with short loops. 
Thus if on such configurations 
an observable does not change under single spin flip,
it is stable within the classical model. 
If in addition it is of the special form $X_C F_x$,
then it is also stable  within the quantum 
model (see section \ref{subsec:q-stable}).

In this section we shall build such observable. 
To this end we shall first define homology classes 
of spin configurations corresponding to 
configurations of loops with short loops only. 
We will then  show that, as expected, single spin flip 
does not change those homology classes. 
This implies that any observable which depends 
solely on the homology classes 
does not change  under single spin flips (for configurations 
containing only short loops). 
This result holds for torus of any dimension.
Subsequently we shall show, that some observables of the 
form $X_C F_x$ share this property.

\subsection{Observables depending only on homology classes}
Let us introduce some notation. 
By $S$ we will denote configuration of spins on the lattice (in the form of configurations of bits
whose values encode spin orientation). 
Given two spin configurations $S_1$ and $S_2$, we can 
add them to obtain new configuration $S$.  We denote it by 
$S=S_1 \oplus S_2$, and the addition is bit-wise, modulo $2$.
I.e. if at given site the spins are the same, resulting spin is down, if they are different, resulting spin is up. 
We denote by $S_0$ configuration of all spins down. 

By $\loops$ we will denote set of excited links. 
A link is excited, if the parity of spins on adjacent
plaquettes is odd (in 3D a link has four such plaquettes, and in 4D -- six ones). One finds that $\loops$ 
is sum of disjoint loops  $l_j$ (the loops can have self crossing
at nodes):

\begin{lemma}
For given configuration $\loops$ consider a connected set of links. 
It is sum of closed loops, which visit 
each link and each node at most one time. 
Equivalently, it is a closed walk, which visit one link 
only at most once. 
\end{lemma}

{\bf Proof}. The proof is by induction.

We will call such connected sets "loops".
We will say that a loop is short, 
when its length is no greater than $cL$,
where $c$ is a fixed constant, which we can take e.g. $1/8$. 

Any spin configuration $S$ defines link configuration $\loops$.
We will then write $S(\loops)$. 
Of course for given $\loops$ there are many spin configurations 
leading to them. Sometimes for given $S$ the corresponding $\loops$ 
will be denoted by $\partial S$ and called boundary of $S$.

\begin{definition}
By continuous deformation of spin configuration 
we mean operation, which can be composed of the following 
elementary operations: flipping spins on all 
plaquettes belonging to an elementary d-dimensional cubes. 
\end{definition}

{\bf Remark 1.} Continuous deformation does not change 
the configuration of links. For 3D easy to see: 
indeed, flipping spins on faces of cube, change at the same time spins on two plaquettes  adjacent to a link from the cube.  


\begin{definition}
\label{def:closed-hom}
We say that $S_1$ and $S_2$ with empty boundary
are homologically equivalent if they can be transformed 
into one another by continuous deformation.
$S$ is called homologically trivial, if it can be continuously transformed into $S_0$.
\end{definition}

\begin{definition}
\label{def:spin-hom}
We say that $S_1$ and $S_2$ which have the same boundary 
are homologically equivalent and denote it by $S_1\sim S_2$, if 
$S_1\oplus S_2$ is homolgically trivial 
\end{definition}

\begin{definition}("Shortest configuration")
\label{def:shortest}
Consider given $\loops=\bigcup_j l_j$.
For each loop $l_j$ fix a shortest surface whose boundary is $l_j$. 
Consider then $S^*(l_j)$  which has spins up on this surface 
and all other spins down. The configuration $S^*=\oplus_j S^*(l_j)$
will be called shortest configuration for $\lcal$.
\end{definition}

\begin{fact}
All shortest configurations $S^*$ for given $\loops$ are homologically equivalent, provided $\loops$ contains only 
short loops. 
\end{fact}

{\bf Proof.}
Take two different shortest configurations. We have 
\be
S_1^*\oplus S_2^* =\oplus_j [S^*_1(l_j)\oplus S^*_2(l_j)]
\ee
However, each configuration 
$S_1^*(l_j)\oplus S_2^*(l_j)$ is trivial. Indeed, since loop $l_j$ 
is short then $|S^*_1(l_j)|$ and $|S^*_2(l_j)|$ are small,
and cannot form homologically nontrivial surface. \blacksquare

\begin{definition}
For $\loops$ containing only short loops, 
with any $S$ leading to $\loops$ 
we can associate  the homology class 
of $S\oplus S^*(\loops)$. 	Denote it by $h(S)$.
\label{def:hom-class-S}
\end{definition}

{\bf Remark 2.} For fixed $\loops$ obviously 
$S_1\sim S_2$  iff $h(S_1)=h(S_2)$. 
Thus the above definition 
allows to ascribe labels to homology classes of 
spin configuration, by relating to distinguished 
class i.e. the class of $S^*$. 
But the homology classes are now defined for any 
$\loops$. Thus we will be able to ask later, 
whether a spin flip (which of course changes $\loops$) 
can preserve homology class.
For any $\loops$ there are eight homology classes in 3D case,
associated with three possible ways of winding around torus. 
In $4D$ there is 16 classes. 


We have obvious fact:
\begin{fact}
\label{fact:sigma-oplus}
We have $S_1 \oplus S_2 = \sigma_j(S_1) \oplus \sigma_j(S_2)$,
where $\sigma_j$ flips $j$-th spin. 
\end{fact}

Now we will show that for short loops, 
single spin flip does not change homology class 
of $S$. To this end we first prove the following
lemma 
\begin{lemma}
\label{lem:star-flip}
For $\loops$ containing only short loops  
we have 
\be
\sigma_i(S^*(\loops))\sim S^*(\sigma_i(\loops)).
\ee
Here $\sigma_i(\loops)$ is understood as the configuration of loops
arising from configuration $\loops$ by applying $\sigma_i$
\end{lemma}
{\bf Proof.} 
Divide $\loops$ into two sets: $\loops_1$
consisting of loops that contain some links from $i$-th plaquette,
and $\loops_2$ which does not contain links from this plaquette. 
Then $\sigma_i(\loops)=\sigma_i(\loops) \cup \loops_1$ hence 
\be
\sigma_i(S^*(\loops))=\sigma_i(S^*(\loops_1))\oplus S^*(\loops_2)
\ee
and 
\be
S^*(\sigma_i(\loops))=S^*(\sigma_i(\loops_1))\oplus S^*(\loops_2).
\ee
Thus only $\loops_1$ is in the game:
\be
\sigma_i(S^*(\loops))\oplus S^*(\sigma_i(\loops))=
\sigma_i(S^*(\loops_1))\oplus S^*(\sigma_i(\loops_1))
\label{eq:proof}
\ee
and therefore we have to show that right-hand-side of the above formula 
is homologically trivial. Indeed, the set $\loops_1$ contains at most two loops independently 
of dimension. Now, since loops are short, 
both $\sigma_i(S^*(\loops_1))$ and  $S^*(\sigma_i(\loops_1))$ are small, 
and added together  must give a trivial surface. 
\blacksquare

Now we are in position to prove the main result of this section
\begin{proposition}
Consider configuration of spins $S$ for which $\loops$ has short loops only.
Then single spin flip does not change the homology class of 
$S$. More explicitly, we have 
\be
\sigma_i(S)\oplus S^*(\sigma_i(\loops))\sim S\oplus S^* (\loops)
\ee
\end{proposition}

{\bf Proof.}
By lemma \ref{lem:star-flip}  we have 
\be
\sigma_i(S^*(\loops))\sim S^*(\sigma_i(\loops)).
\ee
By fact \ref{fact:sigma-oplus} we have 
\be
\sigma_i(S) \oplus \sigma_i( S^*(\loops)= S\oplus S^*(\loops)
\ee
Combining the above two equations, we obtain the claim.\blacksquare

Thus any observable $\Tau$ which for configurations 
$\loops$ containing only short loops depends only 
on homology class, i.e. 
\be
\Tau(S)=\Tau(h)
\ee
is dynamically stable within the model of Proposition 
\ref{prop:cl-model} below some critical temperature. 

\subsection{Construction of stable topological observables}
Our bare observable will be  $X_T=\Pi_{j\in T}\sigma_x^j$ where $T$ is chosen in such a way that $X_T$ is  
invariant under flipping spins on plaquettes from any cube (i.e. 
it is invariant under continuous transformations). Examples of such observables 
exists, as will be shown later in next subsection. 
We will show that one can find dichotomic observable 
$F_x$ which will depend on given configuration $S$ only through $\loops$, 
such that the dressed observable $X_T F_x$ depends only on homology of $S$ 
(for short loops) i.e. it is stable within the classical model. 

We begin with the following lemma
\begin{lemma}
The observable $X_T$ which is invariant under flipping 
spins on plaquettes of any cube is constant on 
homology classes for any fixed link configuration $\loops$ containing only short loops (cf. 
definition \ref{def:hom-class-S}).
\end{lemma}
{\bf Remark.}
Note that this does not mean that $X_T$ is stable.
Indeed, for any {\it fixed} link configuration,
it is constant on the whole homology classes.
However if the link configuration changes, 
it may change sign on the same homology class. 
The stable observable described in previous subsection has the same 
value on a given homology class {\it independently} of 
link configurations, provided there are only short loops. 

{\bf Proof.} 
Consider arbitrary spin configurations $S$ and $S'$
whose boundary is $\loops$, 
and which are in the same class of homology,
i.e. $S_1 \equiv S\oplus S^*$ is homologically equivalent 
to $S_2 \equiv S'\oplus S^*$. Therefore 
$S_1$ can be transformed into $S_2$  by 
flipping spins on a set of elementary cubes.  
This does not change the sign of $X_T$,
so that $X_T$ has the same sign on $S_1$ and $S_2$.
Thus it has the same sign also on $S$ and $S'$. \blacksquare

Now we are in position to build stable observable.
Now let us assume that 
\be
X_T(S_1)X_T(S_2)=X_T(S_1\oplus S_2).
\ee
We stress here that this assumption is easily seen to hold 
for particular observables considered in next subsection. (One can actually show, 
that it is true in general for observables satisfying assumptions of the above lemma). 
Using this we can write 
\be
X_T(S)=X_T(S\oplus S^* \oplus S^*)=X_T(S\oplus S^*)X_T(S^*)
\ee
Since homology class of $S^*$ 
is always the same for short loops (independently on possible amiguity of $S^*$ for given loop) 
then $X_T(S^*)$  depends only on the loops configuration:
$X_T(S^*)=X''(\loops)$, so that $X_T(S)=X_T(S\oplus S^*)X''(\loops)$.
Now, since for fixed loops configuration $X_T$ depends only on homology class and the loops configuration 
for $S\oplus S^*$ is always null (as $S\oplus S^*$ does not 
have a boundary), we get that $X_T(S\oplus S^*)$ depends 
only on homology class of $S\oplus S^*$.
Therefore, according to definition \ref{def:hom-class-S},
it depends only on homology $h$ of $S$. 
Hence  $X_T(S\oplus S^*)=X'(h)$
and we have 
\be
X_T(S)=X'(h) X''(\loops).
\ee
Then the following observable
\be
\Tau(S)=X_T(S) X''(\loops)
\label{eq:tau-xf}
\ee
depends only on $h$.   
The above observable is defined unambiguously  only for spin configurations 
leading to short loops configurations. 
This is because $X''$ is only well defined only on 
short loops configurations. 
We then extend the definition  of $\Tau$ to all spin configurations,
by letting $X''(\loops)=1$ for all 
other loops configurations. Thus we shall take $F_x=X''$ and obtain that $X_T F_x$ depends only on homology of spin configuration, hence is stable within classical model.

\subsection{Observable stable within quantum model} 
\label{subsec:q-stable}
The observable constructed in the previous subsection 
is stable within classical model, because it depends only on 
homology class. However, we know that only special observables 
from the classical model evolve in the same way 
in quantum model. E.g. the  observables of the form 
$X_T F_x$, where $F_x$ is 
from algebra generated by star operators $X_s$,a
and $X_T$ commutes with $Z_c$. Here we shall focus on construction of $X_T$ 
since it determines $F_x$ via considerations of the previous section.

Now, let us note that the first condition 
means simply that $F_x$ depends only on loops.
The second condition means that 
$X_T$ does not change under flips on all plaquettes 
of an elementary cube. Thus the 
observable \eqref{eq:tau-xf} is of the above form,
hence it evolves in the same 
way both in quantum and classical model,
hence it is stable also within quantum model. 

The last thing is to assure that the observable $\Tau$ 
is nontrivial, i.e  it is not identity. 
To this end we have choose the set $T$ in a special way,
such that on spin configurations without boundary,
$X_T$ can take different sign. 


For $3D$ it will be nontrivial loop in dual lattice,
i.e. straight line consisting of parallel plaquettes. 
The fact that it is loop in dual lattice, implies 
that it $X_T$ is invariant under continuous transformations.
Since it is nontrivial, then $X_T$ 
have value $-1$ for spin configuration 
consisting of plane of flipped spins perpendicular to $T$,
while it takes value $1$ on homologically trivial 
spin configurations. Since there are three possible 
choices of inequivalent nontrivial loops,
we can construct three independent observables.

In $4D$ we take $T$ to be plane in dual lattice, i.e. the value observable $X_T$ 
is defined as a product of values of all plaquettes belonging to the plane $T$. 
Again, $X_T$ does not change under flipping spins on cube 
because arbitrary cube has exactly two plaquettes in common with 
such a plane.  
For this reason it will be also $1$ on homologically trivial 
spin configurations. However it will take value $-1$ 
on the configuration consisting of flipped spins on a plane $T'$
whose intersection with  $T$ is a single plaquette. 
Note that since there are six homologically nontrivial planes,
we can construct six independent observables of this sort.  

Now, since the torus in $4D$ is selfdual,
we can consider dual observable i.e. $\Tau_z=Z_T' F_z$, 
and $F_z$ 
depends only on configuration of three dimensional 
cubes (such cubes are dual to link). 
Since $F_z$ and $F_x$ commute, and planes $T$ and $T'$ 
intersect only in a single plaquette, we obtain 
that $\Tau_z$ and $\Tau_x$ anticommute,
so that they form a qubit.

\section{Polynomial algorithm for measuring the topological observables}
\label{sec:alg}
The observables are symmetric, so it is enough to show algorithm 
for one of them, say $\Tau_z$. The algorithm is the following.
\bee
\item Measure all spins. 
\item Multiply outcomes on a fixed plane in dual lattice,
this gives "raw value" of the observable.
\item Identify the loops. 
\item For "short" loops we identify associated surfaces  (the ones homologically equivalent to shortest ones). 
\item If an odd number of surfaces crosses a fixed plane in dual 
lattice, multiply the "raw value" with $-1$. 
\eee

The step 2 corresponds to measuring the bare observable $X_T$, while the steps 3-5 
define observable $F_x$. The multiplication in last step produces the stable, dressed observable $\Tau_x=X_T F_x$.
The only nontrivial problem here is to 
argue that the step $4$ is polynomial.
It is actually enough to show that for a fixed loop, 
one can find efficiently a surface which is contained 
in the smallest cube containing the loop. 

Moreover, it is enough to find a protocol which in efficient way 
allows to find spins which, if flipped, reduce length 
of the loop by some amount (in our protocol, 
it will be reduced by two). 

The protocol is the following. We first choose a Cartesian frame.
We start with a link of the loop, and move along the loop.
If there is ambiguity (the loop crosses itself)
the priority is set by the chosen frame: if only we can we go 
in positive direction of the axis with the smallest number.
If not, then we go in negative direction of the axis with the 
smallest number.  The same rule governs choice of 
the starting link. 

The walk is stopped, if we are forced 
at some point to go in opposite direction to 
any of the previous steps (see figure).

\begin{figure}
  \centering
  \includegraphics[width=7cm]{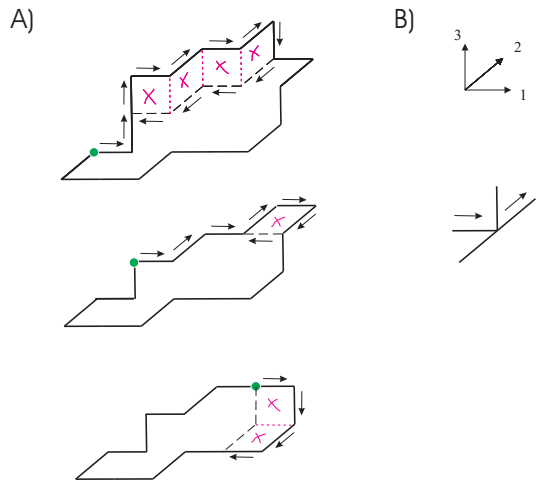}
  \caption{Efficient algorithm for determining surface closing the loop.}
  \label{fig:merging}
\end{figure}

When the walk is stopped, the link at which we stopped 
and the last "opposite" link, determine uniquely 
the set of plaquettes. This is because all the links of the 
walk lying between two "opposite links" 
are perpendicular to them. Now, after flipping spins 
on the set of plaquettes, the two opposite links are removed
from the curve. Note that this flipping may further diminish 
the length of loop, if by a chance, the chose plaquettes 
have some other links common with the loop. 
It may also divide the loop into smaller ones, however 
their joint length  is not longer than $l-2$.

\section{Concluding remarks}

\label{sec:conc}
We have shown that within Markovian weak coupling approximation,
there exist a stable quantum subsystem in four dimensional Kitaev model of
\cite{KitajewPreskill-FTmem}. While the qubit is indeed stable, there are 
several other drawbacks, which  makes the question of existence of self correcting
quantum
memory still open. Minimal requirement for good quantum memory is that it 
should allow to encode arbitrary state of
qubit (encoding), then to store it for long time (storage) and finally
perform a measurement in arbitrary basis (readout). It would be
also good if the measurement
is repeatable. The present result shows that storage is possible, but does
not touch the problem of preparation and measurement.
Actually, the algorithm for measuring topological observables is highly
destructive, hence non-repeatable.
The encoding and read-out one usually performs by preparing qubit in a
standard state, and also measure standard observable, the rest being
done by gates. Also repeatability can be then assured, if one can perform
c-not gates on the protected qubits.
However the problem with the Kitaev's model is that it does not support
universal computation. A possible solution of this problem is to use
the version of topological quantum memory developed by Bombin and Delgado \cite{BombinD2006-top-univ}
which supports universal computation (we shall present the
dynamical analysis of these models elsewhere). However, still there is a
separate problem of preparation of the qubit in standard state.

Finally, let us mention, that  the present result has separate
implications in statistical physics. In the standard approach to large quantum 
systems the metastable states of encoded qubits like those found for
 Kitaev models disappear in the  thermodynamic limit merging into a  
classical simplex  of  equilibrium (KMS) states \cite{AlickiH2006-drive}. On the other hand 
they carry an interesting topological structure which might be physically relevant. 
In this context it is interesting to ask for a new description of infinite system, which would
take into account such new metastable states. In particular, phase
transitions which lead to such curious states require further investigations.

\noindent
\textbf{Acknowledgements} \\
We are grateful to John Preskill for drawing our attention to results of Ref. \cite{KitajewPreskill-FTmem} 
on 4D model and for numerous discussions. We thank Mark Fannes for stimulating discussions. M.H. would like also to thank Hector Bombin and Miguel Martin-Delgado for helpful discussions. This work is supported by the EU Project QAP-IST contract 015848 and EC IP SCALA.

\bibliographystyle{apsrev}
\bibliography{rmp12-hugekey}

\end{document}